\documentclass[a4paper,nobibnotes,nofootinbib,preprintnumbers]{revtex4}
\usepackage{amssymb}
\usepackage{amsmath}
\usepackage{epsfig}
\newcommand{\be}{\begin{equation}}
\newcommand{\ee}{\end{equation}}
\newcommand{\bea}{\begin{eqnarray}}
\newcommand{\eea}{\end{eqnarray}}
\newcommand{\nn}{\nonumber}

\newcommand{\D}{\displaystyle}
\newcommand{\bs}{\boldsymbol}
\newcommand{\g}{\gamma}
\newcommand{\f}{\frac}

\newcommand{\xp}{x_{\mathbb P}}
\newcommand\lr[1]{{\left({#1}\right)}}
\begin{document}
\title{A unified description of diffractive deep inelastic scattering with saturation}
\author{Cyrille Marquet}\email{marquet@quark.phy.bnl.gov}
\affiliation{RIKEN BNL Research Center, Brookhaven National Laboratory, Upton, NY 11973, USA}
\preprint{RBRC-684}
%%%%%%%%%%%%%%%%%%%%%%%%%%%%%%%%%%%%%%%%%%%%%%%%%%%%%%%%%%%%%%%%%%
%%%%%%%%%%%%%%%%%%%%%%%%   Abstract   %%%%%%%%%%%%%%%%%%%%%%%%%%%%
%%%%%%%%%%%%%%%%%%%%%%%%%%%%%%%%%%%%%%%%%%%%%%%%%%%%%%%%%%%%%%%%%%
\begin{abstract}

We propose a new description of inclusive diffraction in deep inelastic scattering
(DIS). The diffractive structure functions are expressed in the dipole picture and
contain heavy-quark contributions. The dipole scattering amplitude, a saturation
model fitted on inclusive DIS data, features a saturation scale $Q_s(x)$ larger than
1 GeV for $x\!=\!10^{-5}.$ The $q\bar qg$ contribution to the diffractive final state
is modeled in such a way that both the large$-Q^2$ and small$-\beta$ limits are
implemented. In the regime $\xp\!<\!0.01$ in which saturation is expected to be
relevant, we obtain a parameter-free description of the HERA data with 
$\chi^2/\mbox{points}\!=\!1.2.$

\end{abstract}
\maketitle
%%%%%%%%%%%%%%%%%%%%%%%%%%%%%%%%%%%%%%%%%%%%%%%%%%%%%%%%%%%%%%%%%%
%%%%%%%%%%%%%%%%%%%%%%   Introduction   %%%%%%%%%%%%%%%%%%%%%%%%%%
%%%%%%%%%%%%%%%%%%%%%%%%%%%%%%%%%%%%%%%%%%%%%%%%%%%%%%%%%%%%%%%%%%
\section{Introduction}

Deep inelastic scattering (DIS) is a process in which a virtual photon is used as a hard probe to resolve the small distances inside a proton and study its partonic constituents: quarks and gluons that obey the laws of perturbative QCD. When probing with a fixed photon virtuality $Q^2\!\gg\!\Lambda_{QCD}^2,$ and increasing the energy of the photon-proton collision $W$, the parton densities seen by the photon inside the proton grow. Eventually, at some energy much bigger than the hard scale, corresponding to a small value of the Bjorken variable $x\!\simeq\!Q^2/W^2,$ the gluon density probed becomes so large that 
non-linear effects like gluon recombination become important. One enters a non-linear yet weakly-coupled regime of QCD \cite{glr} called the saturation regime.

The transition to the saturation regime is characterized by the so-called saturation momentum $Q_s(x)\!=\!Q_0\ x^{-\lambda/2}.$ This is an intrinsic scale of the high-energy proton which increases as $x$ decreases. $Q_0\!\sim\!\Lambda_{QCD},$ but as the energy increases, $Q_s$ becomes a hard scale, and the transition to saturation occurs when $Q_s$ becomes comparable to $Q.$ The higher $Q^2$ is, the smaller $x$ should be to enter the saturation regime. Part of the DIS events are diffractive, meaning that the proton remains intact after the collision and there is a rapidity gap between that proton and the rest of the final-state particles. Such events are expected to be more sensitive to the saturation regime than the inclusive ones.

Although the saturation regime is only reached when $Q_s\!\sim\!Q,$ observables are sensitive to the saturation scale already during the approach to saturation 
\cite{extscal} when $\Lambda_{QCD}\!\ll\!Q_s\!\ll\!Q.$ For inclusive events in deep inelastic scattering, this feature manifests itself via the so-called geometric scaling property: instead of being a function of $Q^2/Q_0^2$ and $x$ separately, the total cross-section is only a function of $\tau\!=\!Q^2/Q_s^2(x),$ up to large values of $\tau.$ Experimental measurements of inclusive DIS are compatible with that prediction \cite{gsinc}. Recently, it was shown \cite{gsdiff} that diffractive observables also feature the geometric scaling behaviors expected when approaching saturation.

In the saturation regime of QCD, contributions to the cross-sections growing like $Q_s/Q$ are important. The leading-twist approximation of perturbative QCD, in which $Q^2$ is taken as the biggest scale, cannot account for such contributions, and therefore is not appropriate to describe the small$-x$ limit of deep inelastic scattering. As leading-twist gluon distributions cannot be used to compute cross-sections, the dipole picture of DIS \cite{dipole} has been developed to describe the high-energy limit. It expresses the hadronic scattering of the virtual photon through its fluctuation into a color singlet 
$q\bar q$ pair (or dipole) of a transverse size $r\!\sim\!1/Q$. The dipole is then the hard probe that resolves the small distances inside the proton. 

The dipole picture naturally incorporates the description of both inclusive and 
diffractive events into a common theoretical framework \cite{nikzak,biapesroy}, as the same dipole scattering amplitudes enter in the formulation of the inclusive and diffractive cross-sections. Different saturation parametrizations of the dipole-proton cross-section have been successful in describing inclusive and diffractive HERA data. The pioneering work of \cite{gbwinc,gbwdiff} triggered several improvements: the "DGLAP-improved" model of \cite{bgbk} allows to include even 
high$-Q^2$ data in the fit and the "BK-inspired" model of \cite{iim} incorporates features from the QCD non-linear equations.

In diffractive DIS, when the invariant mass $M_X$ of the diffractive final-state is much smaller than $Q,$ the dominant contribution to the final state comes from the $q\bar q$ component of the photon wavefunction. By contrast, if $\beta\!\simeq\!\!Q^2/M_X^2\!\ll\!1,$ then the dominant contributions come from the $q\bar qg$ component, or from higher Fock states, i.e. from the photon dissociation. The main goal of this work is to improve the description of the 
$q\bar qg$ contribution with respect to previous analysis: it will be modeled in such a way that both the large$-Q^2$ and small$-\beta$ limits are implemented.

Including the contributions of heavy quarks in the models has also been a recent concern, as several aproaches observed a decrease of the saturation scale to
$Q_s\!\sim\!\Lambda_{QCD},$ when trying to include the charm quark in their analysis \cite{seb,kmm}. This was problematic, however it was recently shown \cite{greg} that it is possible to accomodate the model of \cite{iim} with heavy-quark contributions and a saturation scale that stays above 1 GeV for $x\!=\!10^{-5},$ rather than dropping to about 500 MeV as is the case in other studies. Our second goal in this paper is to check whether the dipole cross-section of this heavy-quark improved saturation model also describes the inclusive diffraction data from HERA.

The plan of the paper is as follows. In Section II, we recall the QCD dipole picture for inclusive and diffractive DIS in terms of the dipole-proton scattering. In Section III, we discuss in more details the case of diffraction and present the different components of the model, highlighting in each case the improvements with respect to previous approaches, in particular concerning the inclusion of heavy-quark contributions, and the treatment of impact parameter. Section IV discusses how to implement the $q\bar qg$ contribution to the diffrative final state to obtain a unified description that features both the large$-Q^2$ and small$-\beta$ limits.
In Section V, the results of the comparison with the available HERA data are presented, and Section VI is devoted to conclusions.

%%%%%%%%%%%%%%%%%%%%%%%%%%%%%%%%%%%%%%%%%%%%%%%%%%%%%%%%%%%%%%%%%%
%%%%%%%%%%%%%%%%%%%%%%%%   Section 1   %%%%%%%%%%%%%%%%%%%%%%%%%%%
%%%%%%%%%%%%%%%%%%%%%%%%%%%%%%%%%%%%%%%%%%%%%%%%%%%%%%%%%%%%%%%%%%
\section{The QCD dipole picture of deep inelastic scattering}

\begin{figure}[t]
\begin{minipage}[t]{85mm}
\vspace{-4.7cm}\centerline{\epsfxsize=8.0cm\epsfbox{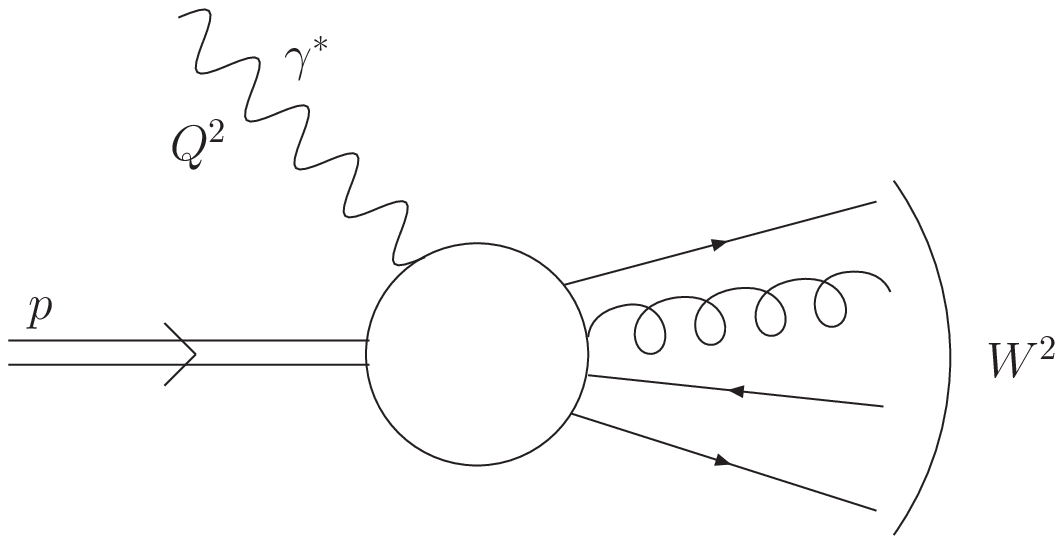}}
\end{minipage}
\hspace{\fill}
\begin{minipage}[t]{85mm}
\centerline{\epsfxsize=7.5cm\epsfbox{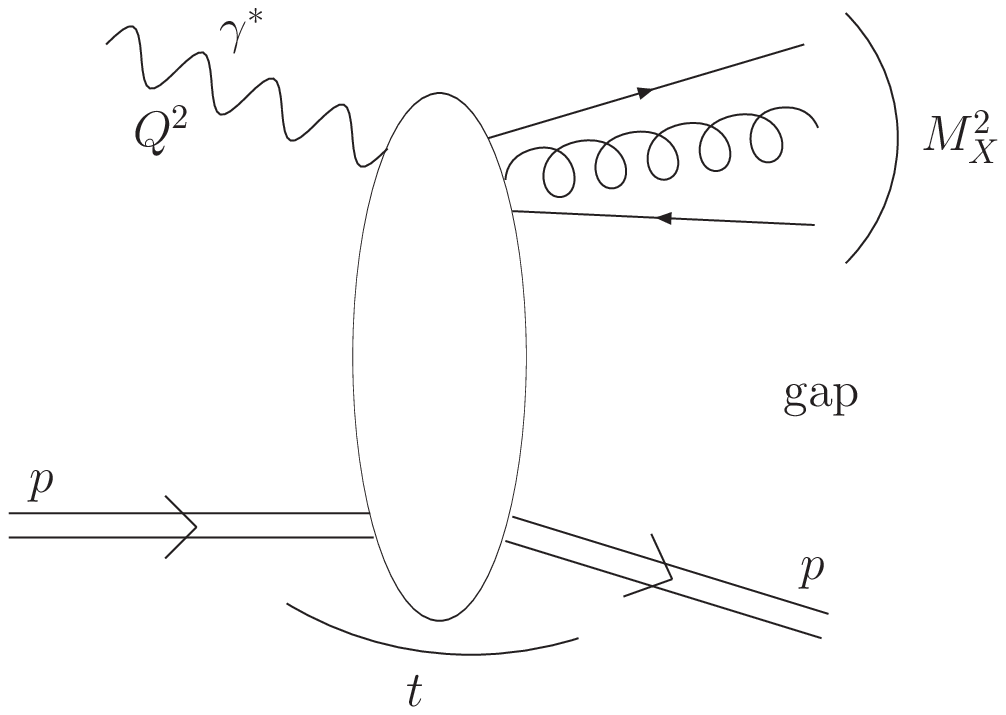}}
\end{minipage}
\caption{Representation of $\g^*\!-\!p$ deep inelastic scattering; inclusive (left) and diffractive (right) events are pictured with the relevant kinematic variables: the photon virtuality $Q^2,$ the energy squared of the $\g^*\!-\!p$ collision $W^2,$ and in the case of diffraction the momentum transfer $t$ and the invariant mass of the diffractive final state $M_X^2.$}
\end{figure}

We focus on diffractive DIS: $\g^*p\!\rightarrow\!Xp$ (see Fig.1). With a momentum transfer 
$t\!\leq\!0,$ the proton gets out of the $\g^*\!-\!p$ collision intact, and there is a rapidity gap between that proton and the final state $X$ whose invariant mass we denote $M_X.$ We recall that the photon virtuality is denoted $Q^2,$ and the $\g^*\!-\!p$ total energy $W.$ It is convenient to introduce the following variables:
\be
x=\f{Q^2}{Q^2+W^2}\ ,\hspace{1cm}\beta=\f{Q^2}{Q^2+M_X^2}\ ,\hspace{1cm}\xp=x/\beta\ .\ee
The $\g^*\!-\!p$ total cross-section $\sigma^{\g^*p\rightarrow X}_{tot}$ is 
usually expressed as a function of $x$ and $Q^2,$ while the diffractive 
cross-section $d\sigma^{\g^*p\rightarrow Xp}_{diff}/d\beta dt$ is expressed as a function of $\beta,$ $\xp,$ $Q^2,$ and $t.$ The size of the rapidity gap in the final state is $\ln(1/\xp).$

\subsection{The $\g^*\to q\bar q$ wavefunctions}

To compute those cross-sections in the high-energy limit, it is convenient to 
view the process in a particular frame called the dipole frame. In this frame, 
the virtual photon undergoes the hadronic interaction via a fluctuation into a 
colorless $q\bar q$ pair, called dipole, which then interacts with the target 
proton. The wavefunctions $\psi_\lambda^{f,\alpha\beta}(z,\textbf{r};Q^2)$ 
describing the splitting of a virtual photon with polarization $\lambda$ into a 
dipole are well known. The indices $\alpha$ and $\beta$ denote the spins of the 
quark and the antiquark composing the dipole of flavor $f.$ The wavefunctions 
depend on $Q^2,$ the fraction $z$ of longitudinal momentum (with respect to the 
$\g^*\!-\!p$ collision axis) carried by the quark, and the two-dimensional 
vector $\textbf{r}$ whose modulus is the transverse size of the dipole.

Formulae giving the functions $\psi_\lambda^{f,\alpha\beta}$ can be 
found in the literature (see for instance~\cite{kovmc}). In what follows, we will need the functions $\Phi^f_\lambda$ which describe the overlap between two wavefunctions for splitting into dipoles of different transverse size 
$\textbf{r}$ and $\textbf{r}':$
\be
\phi^f_\lambda(z,\textbf{r},\textbf{r}';Q^2)=N_c\sum_{\alpha\beta}
\left[\psi_\lambda^{f,\alpha\beta}(z,\textbf{r}';Q^2)\right]^*
\psi_\lambda^{f,\alpha\beta}(z,\textbf{r};Q^2)\ .\label{overlap}
\ee
For a transversely (T) or longitudinally (L) polarized photon, these functions 
are given by
\be
\phi^f_T(z,\textbf{r},\textbf{r}';Q^2)=
\frac{\alpha_{em}N_c}{2\pi^2}e_f^2
\left((z^2+(1\!-\!z)^2)\varepsilon_f^2
\f{\textbf{r}.\textbf{r}'}{|\textbf{r}||\textbf{r}'|}
K_1(\varepsilon_f|\textbf{r}|)K_1(\varepsilon_f|\textbf{r}'|)
+m_f^2 K_0(\varepsilon_f|\textbf{r}|)
K_0(\varepsilon_f|\textbf{r}'|)\right)\ ,\label{trans}\ee
\be
\phi^f_L(z,\textbf{r},\textbf{r}';Q^2)=
\frac{\alpha_{em}N_c}{2\pi^2}e_f^2
4Q^2 z^2(1\!-\!z)^2 K_0(\varepsilon_f|\textbf{r}|)
K_0(\varepsilon_f|\textbf{r}'|)\ .\label{long}\ee
In the above, $e_f$ and $m_f$ denote the charge and mass of the quark with 
flavor $f$ and 
\be
\varepsilon_f^2\!=\!z(1\!-\!z)Q^2\!+\!m_f^2\ .
\ee

\begin{figure}[t]
\begin{minipage}[t]{60mm}
\centerline{\epsfxsize=6cm\epsfbox{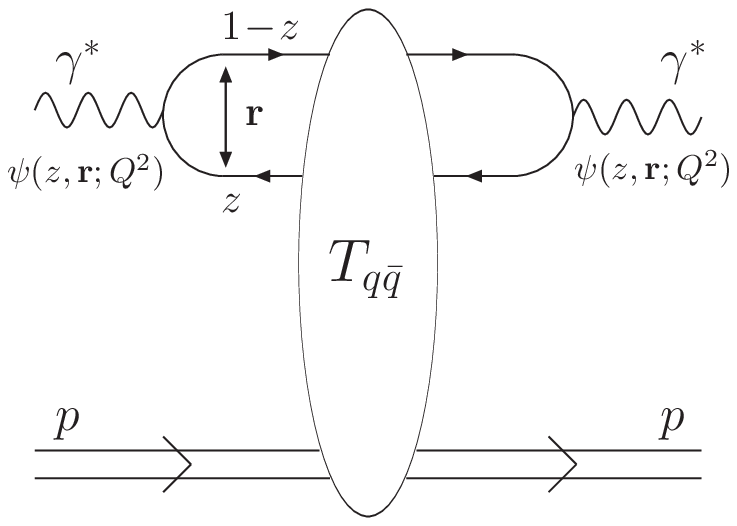}}
\end{minipage}
\hspace{\fill}
\begin{minipage}[t]{110mm}
\centerline{\epsfxsize=11cm\epsfbox{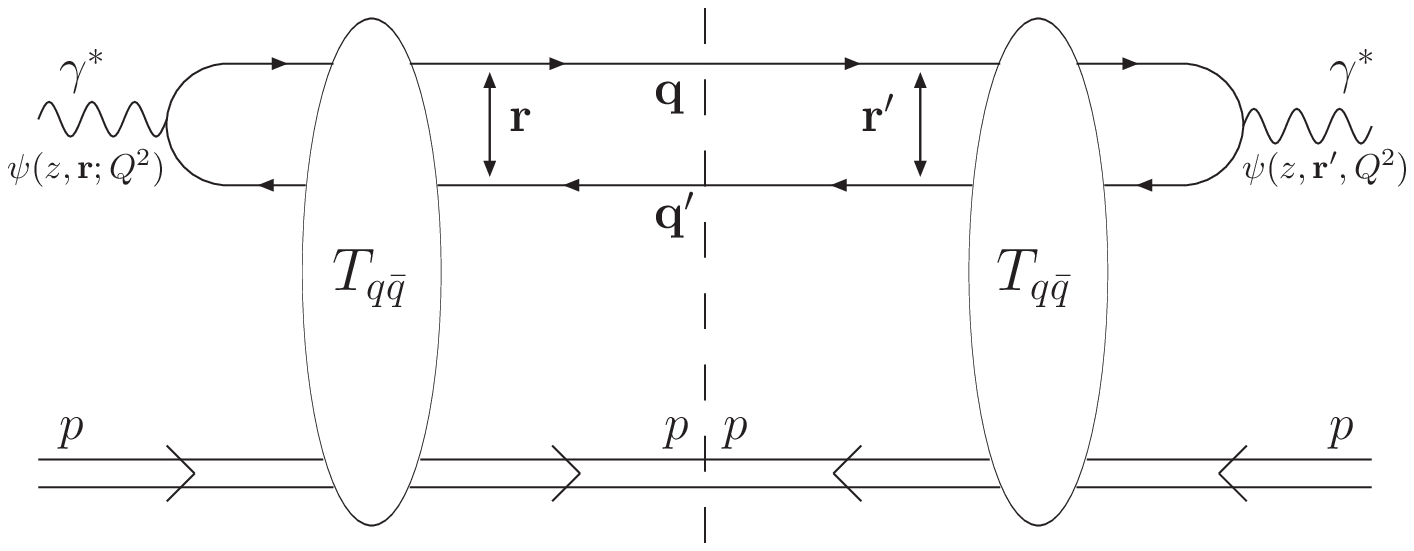}}
\end{minipage}
\caption{The QCD dipole picture of deep inelastic scattering. The left diagram represents $\g^*\!-\!p$ elastic scattering and (via the optical theorem) corresponds to formula \eqref{tot}. The right diagram represents diffractive scattering (without possible final states containing gluons) and corresponds to
formula \eqref{diff}. In this case, the final state (indicated by the vertical dashed line) is characterized by $t\!=\!-\bs{\Delta}^2$ and 
$M_X^2\!=\!(\bs{\kappa}^2\!+\!m_f^2)/(z(1-z)),$ with 
$\bs{\Delta}\!=\!\textbf{q}\!+\!\textbf{q}'$ and 
$\bs{\kappa}\!=\!(1\!-\!z)\textbf{q}\!-\!z\textbf{q}'$ in terms of the quark and antiquark momenta $\textbf{q}$ and $\textbf{q}'.$ Via Fourier transformations, $\textbf{q}$ and 
$\textbf{q}'$ impose different sizes and impact parameters for the dipole in the amplitude and the dipole in the complex conjugate amplitude.}
\end{figure}

\subsection{The total cross-section $\sigma^{\g^*p\rightarrow X}_{tot}$} 

Via the optical theorem, the $\g^*\!-\!p$ total cross-section is related to the elastic scattering of the virtual photon off the proton. In the dipole frame, this happens as follows: at a given impact parameter $\textbf{b},$ the photon splits into a dipole with a given size $\textbf{r}$ which scatters elastically off the proton and recombines back into the photon. Therefore the overlap function $\Phi_\lambda$ which enters in the computation of the total cross-section is
\be
\Phi_\lambda(z,|\textbf{r}|;Q^2)=\sum_f
\phi^f_\lambda(z,\textbf{r},\textbf{r};Q^2)\ .
\ee
For a virtual photon with polarization $\lambda,$ the total cross-section is then given by (see Fig.2a):
\be
\sigma^{\g^*p\rightarrow X}_\lambda(x,Q^2)=2
\int d^2r \int_0^1 dz\ \Phi_\lambda^{\g^*\g^*}(z,|\textbf{r}|;Q^2)\
\int d^2b\ T_{q\bar q}(\textbf{r},\textbf{b};x)\label{tot}\ee
where the function $T_{q\bar q}(\textbf{r},\textbf{b};x)$ is the elastic 
scattering amplitude of the dipole of size $\textbf{r}$ off the proton at impact 
parameter $\textbf{b}.$ It contains the $x$ dependence, reflecting the fact that 
in our frame, the proton carries all the energy and is therefore evolved up to 
the rapidity $\ln(1/x).$ In the high-energy limit $x\!\ll\!1$ we are considering 
here, $T_{q\bar q}$ does not depend on $z.$

\subsection{The diffractive cross-section 
$d\sigma^{\g^*p\rightarrow Xp}_{diff}/d\beta dt$}

The diffractive scattering happens as follows. In the amplitude, the photon splits into a dipole of size $\textbf{r}$ which scatters off the proton at a given impact parameter 
$\textbf{b}$ and dissociates into a final state of invariant mass $M_X.$ The same happens in the complex conjugate amplitude, except that the dipole size 
$\textbf{r}'$ and the impact parameter $\textbf{b}'$ are different from $\textbf{r}$ and 
$\textbf{b}.$ Indeed, the final state is characterized by particular values of 
$M_X$ (or equivalently $\beta$) and $t,$ corresponding to particular momenta of the quark and antiquark in the final state. In coordinate space, this imposes two different dipole sizes and impact parameters in the amplitude and the complex conjugate amplitude, therefore the functions 
$\phi_\lambda^f(z,\textbf{r},\textbf{r}';Q^2)$ (see \eqref{overlap}) 
enter in the computation of the diffractive cross-section. For a virtual photon with polarization $\lambda,$ the diffractive cross-section is given by (see Fig.2b):
\bea
\f{d\sigma^{\g^*p\rightarrow Xp}_\lambda}{d\beta\ dt}(\beta,\xp,Q^2,t)=
\frac{Q^2}{4\beta^2}\sum_f \int \f{d^2r}{2\pi} \int \f{d^2r'}{2\pi}
\int_0^1 dz z(1\!-\!z) \Theta(\bs{\kappa}_f^2)\
e^{i\bs{\kappa}_f.(\textbf{r}'-\textbf{r})}\hspace{3cm}\nn\\
\phi_\lambda^f(z,\textbf{r},\textbf{r}';Q^2)
\int d^2b\ d^2b'\ e^{i\bs{\Delta}.(\textbf{b}'-\textbf{b})} 
T_{q\bar q}(\textbf{r},\textbf{b};\xp)
T_{q\bar q}(\textbf{r}',\textbf{b}';\xp)\ .\label{diff}\eea
In the above, the differences between $\textbf{r}$ and $\textbf{r}'$ on one hand, and $\textbf{b}$ and $\textbf{b}'$ on the other hand, are related via Fourier transformation to
\be
\bs{\kappa}_f^2\!=\!z(1\!-\!z)Q^2(1\!-\!\beta)/\beta\!-\!m_f^2\ ,\hspace{0.5cm}
\mbox{ and }\hspace{0.5cm}\bs{\Delta}^2=-t\ .\ee
Note that now, the proton is only evolved up to the rapidity $\ln(1/\xp).$ This 
is because some of the energy ($M_X^2$) is carried by the dipole in order to 
form the diffractive final state. The dipole is evolved up to a rapidity 
$\ln(1/\beta)$ and the proton up to the rapidity $\ln(\beta/x)\!=\!\ln(1/\xp).$ 
The relevant high-energy limit in this case is $\xp\!\ll\!1.$

Note that to write formula \eqref{diff}, we have neglected possible final states containing gluons. This is justified because these are suppressed by extra powers of 
$\alpha_s.$ However, if $\beta$ becomes too small, or if $Q^2$ becomes too large, the dipole will emit soft or collinear gluons whose emissions are accomponied by large logarithms $\ln(1/\beta)$ or $\ln(Q^2)$ which will compensate the factors of 
$\alpha_s.$ This will be discussed in more details in Section IV, when we explain how to implement the $q\bar qg$ contribution to the diffractive final state, in order to correctly describe both the small$-\beta$ and large$-Q^2$ limits.

%%%%%%%%%%%%%%%%%%%%%%%%%%%%%%%%%%%%%%%%%%%%%%%%%%%%%%%%%%%%%%%%%%
%%%%%%%%%%%%%%%%%%%%%%%%   Section 2   %%%%%%%%%%%%%%%%%%%%%%%%%%%
%%%%%%%%%%%%%%%%%%%%%%%%%%%%%%%%%%%%%%%%%%%%%%%%%%%%%%%%%%%%%%%%%%
\section{The saturation model for the dipole amplitude $T_{q\bar q}$}

Using the dipole picture of deep inelastic scattering, we have expressed the total 
\eqref{tot} and diffractive \eqref{diff} cross-sections in the high-energy limit in terms of a single object: the dipole scattering amplitude off the proton 
$T_{q\bar q}(\textbf{r},\textbf{b};x).$ It is mainly a non-perturbative quantity, but its evolution towards small values of $x$ (or high energy) is computable from perturbative QCD. Evolution equations have been established in the leading 
$\ln(1/x)$ approximation \cite{bk,jimwlk,ploop} and, at least for central impact parameters, one has learned a lot about the growth of the dipole amplitude and the transition from the leading-twist regime $T_{q\bar q}\!\ll\!1$ towards and into the saturation regime $T_{q\bar q}\!\lesssim\!1.$

Let us recall that this transition is characterized by the saturation scale 
$Q_s(x),$ which increases as $x$ decreases. In the following, we shall work in the context of the Balitsky-Kovchegov (BK) evolution \cite{bk} to describe the 
$\textbf{r}$ dependence. Indeed, this provides a natural explanation for the geometric scaling properties of the data \cite{gsinc,gsdiff}. The impact parameter dependence of $T_{q\bar q}$ is still an open problem, it cannot be extracted from perturbative QCD and it is usually modeled. In formulae, one writes
\be
T_{q\bar q}(\textbf{r},\textbf{b};x)=S(\textbf{b})\ N(|\textbf{r}|Q_s(x),x)
\label{bfact}\ee
where we have introduced the factorized impact-parameter profile $S(\textbf{b}).$ In the following, we detail the different components of our model: 
$S(\textbf{b})$ and $N(|\textbf{r}|Q_s,x).$

\subsection{The impact-parameter profile $S(\textbf{b})$}

When performing the $\textbf{b}$ integration in formula \eqref{tot}, this contributes only to the normalization via a constant factor $2\int d^2b\ S(\textbf{b})\!=\!\sigma_0,$ (of order 25 mb) caracterising the transverse area of the proton. However, in the case of the diffractive cross-section \eqref{diff}, the $\textbf{b}$ integration gives the momentum transfer dependence. Experimentally, the diffractive cross-section decreases exponentially with $|t|$ as $e^{B_D t},$ where $B_D$ is the diffractive slope (of order 
$6\ \mbox{GeV}^{-2}$). This is consistent with the Gaussian profile 
$S(\textbf{b})\!=\!e^{-\textbf{b}^2/(2B_D)},$ which then implies 
$\sigma_0\!=\!4\pi B_D\!.$

In the literature, the quantities $\sigma_0$ and $B_D$ are usually considered unrelated, however as we have shown, a consistent treatment of the impact parameter dependence within the dipole picture implies that this is not the case. To summarize, one has:
\be
\f{d\sigma}{dt}\sim e^{B_D t}\hspace{0.5cm}\Rightarrow\hspace{0.5cm}
S(\textbf{b})=e^{-\textbf{b}^2/(2B_D)}\hspace{0.5cm}\Rightarrow
\hspace{0.5cm}\sigma_0=4\pi B_D\label{consipd}\ .
\ee

\subsection{The heavy-quark improved IIM saturation model for $N(|\textbf{r}|Q_s,x)$}

The Iancu-Itakura-Munier (IIM) saturation model is inspired by universal properties
\cite{tw} of solutions of the BK equation \cite{bk}. The most important feature is probably the geometric scaling regime: at small values of $x,$ instead of being a function of a priori the two variables $r\!=\!|\textbf{r}|$ and $x,$ $N$ is actually a function of the single variable 
$rQ_s(x)$ up to inverse dipole sizes significantly larger than the saturation scale $Q_s(x).$ If $rQ_s\!>\!1$ then $N\!=\!1$ and the scaling is obvious. We insist that the scaling property is a non-trivial prediction for $rQ_s\!\ll\!1,$ when $N$ is still much smaller than 1. 

Of course the geometric scaling window has a limited extension: at very small dipole sizes, deep into the leading-twist regime, the scaling breaks down. Universal scaling violations \cite{tw} due to $x$ not being small enough have also been derived and are implemented in the IIM model, which is therefore a function of 
$rQ_s$ and $x.$ Recently, a new type of geometric scaling violations has been predicted, due to the inclusion of Pomeron loops in the evolution 
\cite{ploop,diffscal} (the BK equation only resums fan diagrams). These violations transform the geometric scaling regime into an intermediate energy regime, as they arise at very small values of $x$ in the so-called diffusive scaling regime. This new regime is likely out of the reach of HERA and we shall not address it in this study.

In the IIM model, the saturation scale is parametrized by
\be
Q_s(x)=\lr{\f{x_0}{x}}^{\f{\lambda}2}\ \mbox{GeV}
\ee
and the dipole amplitude is given by
\be
N(rQ_s,x)=\left\{\begin{array}{lll}N_0\ \lr{\f{rQ_s}2}^{2\g_c}\
\exp\left[-\f{2\ln^2(rQ_s/2)}{\kappa\lambda\ln(1/x)}\right]
&\mbox{for }rQ_s\leq 2\\\\1-e^{-4\alpha\ln^2(\beta rQ_s)}
&\mbox{for }rQ_s>2\end{array}\right.
\label{dipcgc}\ee
with $\alpha$ and $\beta$ uniquely determined from the conditions that $N$ 
and its derivative are continuous at $rQ_s\!=\!2.$ The amplitude at the matching point is chosen to be $N_0=0.7.$

In this work, we shall consider the IIM saturation model \cite{iim} extended in
\cite{greg} to include heavy quarks (with $m_c\!=\!1.4\ \mbox{GeV},$ 
$m_b\!=\!4.5\ \mbox{GeV},$ and $m_f\!=\!0.14\ \mbox{GeV}$ for the light flavors). The coefficient $\kappa\!=\!9.9$ is obtained from the BFKL kernel while the critical exponent
$\g_c\!=\!0.7376$ is fitted to the HERA measurements of the proton structure function, along with the remaining parameters. The saturation scale parameters are $\lambda\!=\!0.2197$ and $x_0\!=\!1.632\ 10^{-5}$ and the cross-section at saturation is 
$\sigma_0\!=\!70.26\ \mbox{GeV}^{-2}$ (or $27.36\ \mbox{mb}$). Note that, via 
$\sigma_0\!=\!4\pi B_D,$ this corresponds to the diffractive slope 
$B_D\!=\!5.591\ \mbox{GeV}^{-2},$ which is in agreement with the experimental observations \cite{zeuslps,h1fps}.

\subsection{The $q\bar q$ components of the diffractive structure functions}

Let us introduce the transverse and longitudinal diffractive structure functions
$F_T^{D,3}(\beta,\xp,Q^2)$ and $F_L^{D,3}(\beta,\xp,Q^2).$ They are easily obtained from the diffractive cross-sections 
$d\sigma^{\g^*p\rightarrow Xp}_\lambda/d\beta,$ integrated over the momentum transfer $t.$ In practice, one does not actually carry out the $t$ integration of \eqref{diff}, but one rather uses the fact that the diffractive cross-section decreases exponentially with $|t|$ like $e^{B_D t}.$ One writes:
\be
\xp F_\lambda^{D,3}=\f{Q^2\beta}{4\pi^2\alpha_{em}}
\f{d\sigma^{\g^*p\rightarrow Xp}_\lambda}{d\beta}\ ,\hspace{0.5cm}
\f{d\sigma^{\g^*p\rightarrow Xp}_\lambda}{d\beta}=
\int_{t_{min}}^0 dt\f{d\sigma^{\g^*p\rightarrow Xp}_\lambda}{d\beta dt}\simeq\f1{B_D}
\left.\f{d\sigma^{\g^*p\rightarrow Xp}_\lambda}{d\beta dt}\right|_{t=0}\ ,
\label{cstosf}\ee
with $e^{B_D t_{min}}\!\ll\!1$ (in practice, $t_{min}\!=\!-1\ \mbox{GeV}^2$). When computing \eqref{diff} for $t\!=\!0,$ the two impact parameter integrations yield the factor $\sigma_0^2/4.$ As already discussed, a consistent treatment of the impact parameter dependence of the dipole scattering amplitude
$T_{q\bar q}$ implies $\sigma_0^2/(4B_D)\!=\!\pi\sigma_0,$ which we shall use in what follows.

Note that one could also study $F_T^{D,4}$ and $F_L^{D,4}$ directly. However, there is less data for those $t-$dependent structure functions, and they would not further test our model, which has the exponential decrease $e^{B_D t}$ built in. This type of measurement would rather be interesting to test saturation models which feature a $t-$dependent saturation scale, as predicted in \cite{nfgs} from the full BK equation.

Let us come back to the diffractive structure functions $F_\lambda^{D,3}.$ From formulae \eqref{cstosf}, \eqref{diff} and \eqref{bfact}, one obtains the contributions from the $q\bar q$ final state. Using the transverse overlap function \eqref{trans}, one gets
\bea
\xp F_T^{q\bar q}(\beta,\xp,Q^2)=
\f{\sigma_0 N_c}{32\pi^3}\f{Q^4}{\beta}\sum_f e_f^2\int_0^1 dz\ \Theta(\kappa_f^2) 
z(1\!-\!z)\left[(z^2+(1\!-\!z)^2)(z(1\!-\!z)Q^2\!+\!m_f^2)I_1^2(\kappa_f,\epsilon_f,Q_s)
\right.\nn\\\left.+m_f^2I_0^2(\kappa_f,\epsilon_f,Q_s)\right]
\label{qqT}\eea
for the $q\bar q$ contribution to the transverse diffractive structure function. With the longitudinal overlap function \eqref{long}, one gets the
$q\bar q$ contribution to the longitudinal diffractive structure function:
\be 
\xp F_L^{q\bar q}(\beta,\xp,Q^2)=
\f{\sigma_0 N_c}{32\pi^3}\f{Q^4}{\beta}\sum_f e_f^2\int_0^1 dz
\Theta(\kappa_f^2) 4Q^2z^3(1\!-\!z)^3I_0^2(\kappa_f,\epsilon_f,Q_s)\ .
\label{qqL}\ee
In \eqref{qqT} and \eqref{qqL}, the functions $I_\lambda$ are given by
\be
I_\lambda(\kappa,\epsilon,Q_s)=\int_0^\infty rdr J_\lambda(\kappa r)
K_\lambda(\epsilon r)N(rQ_s,\xp)
\ee
in terms of the dipole scattering amplitude $N(rQ_s,\xp)$ and of the Bessel functions 
$J_\lambda$ and $K_\lambda.$

%%%%%%%%%%%%%%%%%%%%%%%%%%%%%%%%%%%%%%%%%%%%%%%%%%%%%%%%%%%%%%%%%%
%%%%%%%%%%%%%%%%%%%%%%%%   Section 3   %%%%%%%%%%%%%%%%%%%%%%%%%%%
%%%%%%%%%%%%%%%%%%%%%%%%%%%%%%%%%%%%%%%%%%%%%%%%%%%%%%%%%%%%%%%%%%
\section{The $q\bar qg$ contribution to the diffractive final state}

As pictured in Fig.2, formula \eqref{diff} is the contribution of the $q\bar q$ final state
to the diffractive cross-section. We have neglected possible final states containing gluons, and in general it is justified because these are suppressed by extra powers of 
$\alpha_s.$ However, there are two kinematical regimes for which this is not the case: the large$-Q^2$ limit and the small$-\beta$ limit. In those situations, gluon emissions are accomponied by large logarithms $\ln(Q^2)$ or $\ln(1/\beta)$ which compensate the factors of $\alpha_s,$ and multiple gluons emissions should be resummed to complete formula 
\eqref{diff}.

In practice, including the $q\bar qg$ final state is enough to describe the HERA data, and this can be done within the dipole picture in both limits, at leading $\ln(Q^2)$ 
\cite{gwus,gbwdiff} or leading $\ln(1/\beta)$ accuracy \cite{gbar,gkop,gkov,gmun,gmar}, as we recall in this section. Note that, at leading $\ln(1/\beta)$ accuracy, all multiple soft gluon emissions can also accounted for in the dipole picture \cite{kovlev,diffscal}, but we shall restric this phenomenological study to the $q\bar qg$ contribution.

The most popular approach is to consider the large$-Q^2$ limit to implement the $q\bar qg$ contribution \cite{gbwdiff,fss,seb}, even though the experimental measurements do not reach very high values of $Q^2.$ In fact, the contribution of the $q\bar qg$ final state is important only for small values of $\beta$ which, due to the finite energy available, correspond to rather small values of $Q^2.$ This is not satisfactory. In this paper, the 
$q\bar qg$ contribution to the diffractive final state is modeled in such a way that both the large$-Q^2$ and small$-\beta$ limits are implemented.

\subsection{The large$-Q^2$ limit}

At large $Q^2,$ the contribution of the $X\!=\!q\bar qg$ final state in diffractive 
$\g^*p\rightarrow Xp$ scattering was computed in \cite{levwus,gwus}. In momentum space, the collinear gluon has a transverse momentum much smaller than $Q^2.$ In coordinate space, the scattering involves a gluonic $gg$ dipole (see Fig.3a): the transverse distance between the quark and the antiquark is much smaller than the transverse distance between the quark and the gluon. The $q\bar q$ pair on one side and the gluon on the other side form an effective gluonic color dipole which undergoes the hadronic interaction \cite{gwus}. We shall denote the corresponding scattering amplitude off the proton 
$T_{gg}(\textbf{r},\textbf{b};x)$ for a dipole of size $\textbf{r}$ at impact parameter 
$\textbf{b}.$ With our model for impact parameter dependence, we write
\be
T_{gg}(\textbf{r},\textbf{b};\xp)=S(\textbf{b})\ 
\tilde{N}(|\textbf{r}|Q_s(x),\xp)
\ee
where $\tilde{N}$ is the equivalent of $N$ but for a $gg$ dipole.

At leading $\ln(Q^2),$ the $q\bar qg$ final state contributes only to the transverse diffractive structure function and one has
\be
\xp F_T^{q\bar qg}|_{LL(Q^2)}(\beta,\xp,Q^2)=\f{\sigma_0\alpha_s C_FN_c\beta}{32\pi^4}\sum_f e_f^2\int_0^{Q^2}dk^2\ln\lr{\f{Q^2}{k^2}}\int_\beta^1 dz
\left[\lr{1\!-\!\f{\beta}z}^2+\lr{\f{\beta}z}^2\right]I_g^2(\sqrt{1\!-\!z},\sqrt{z},Q_s/k)
\label{qqgQ2}\ee
with
\be
I_g(a,b,c)=\int_0^\infty rdr J_2(ar)K_2(br)\tilde{N}(cr,\xp)\ .
\label{intj2k2}\ee
The computation of \cite{gwus} is a leading-twist two-gluon exchange calculation in which the $gg$ dipole is given by $\tilde{N}(rQ_s,x)\!=\!N_c\ N(rQ_s,x)/C_F$ in terms of the 
$q\bar q$ dipole. However this is not consistent with the use of a saturation model. For instance, $\tilde{N}$ should saturate at 1, not at $N_c/C_F.$ This implies that, when using \eqref{qqgQ2}\!-\!\eqref{intj2k2} with $\tilde{N}\!=\!N_c\ N/C_F$ and a saturation model for $N,$ the analysis in the literature overestimate the $q\bar qg$ contribution. In practice, this is usually compensated by using an unphysically small value for $\alpha_s.$

The parametrization we shall use in this paper is $\tilde{N}\!=\!2N\!-\!N^2.$ This relation implies the large$-N_c$ limit, and therefore goes well with our model for the 
$q\bar q$ dipole scattering amplitude $N:$ it is consistent with the BK evolution implemented in \eqref{dipcgc}. Numerically, this reduces the $q\bar q g$ contribution (with respect to using $\tilde{N}\!=\!N_c\ N/C_F$), especially because the saturation scale is quite large, and therefore $N$ is not always small.

Finally, when computing the heavy quark contributions $c\bar c g$ and $b\bar b g,$ we replace the $\beta$ variable in \eqref{qqgQ2} by $\beta(1\!+\!4m_f^2/Q^2).$ This substitution, which modifies only the small$-Q^2$ results, is necessary in order to insure that there is no $q\bar qg$ contributions when the final state is such that $M_X\!=\!2m_f$ (in practice, such a substitution doesn't make a difference for the light quarks). 

\begin{figure}[t]
\begin{minipage}[t]{55mm}
\centerline{\epsfxsize=5.5cm\epsfbox{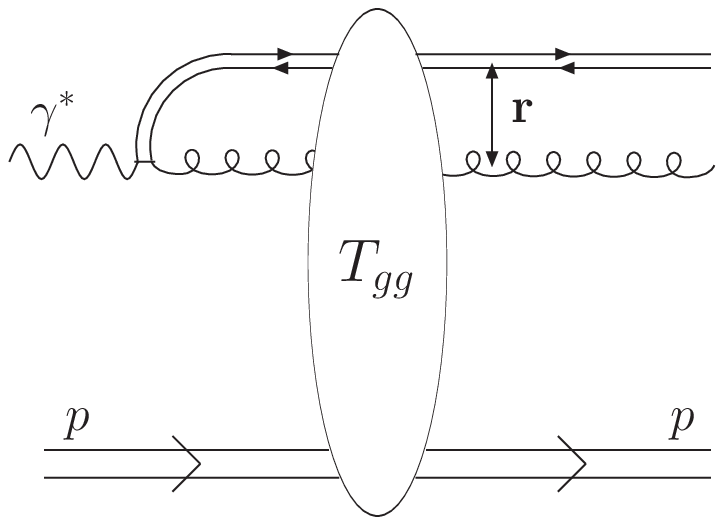}}
\end{minipage}
\hspace{\fill}
\begin{minipage}[t]{114mm}
\centerline{\epsfxsize=11.4cm\epsfbox{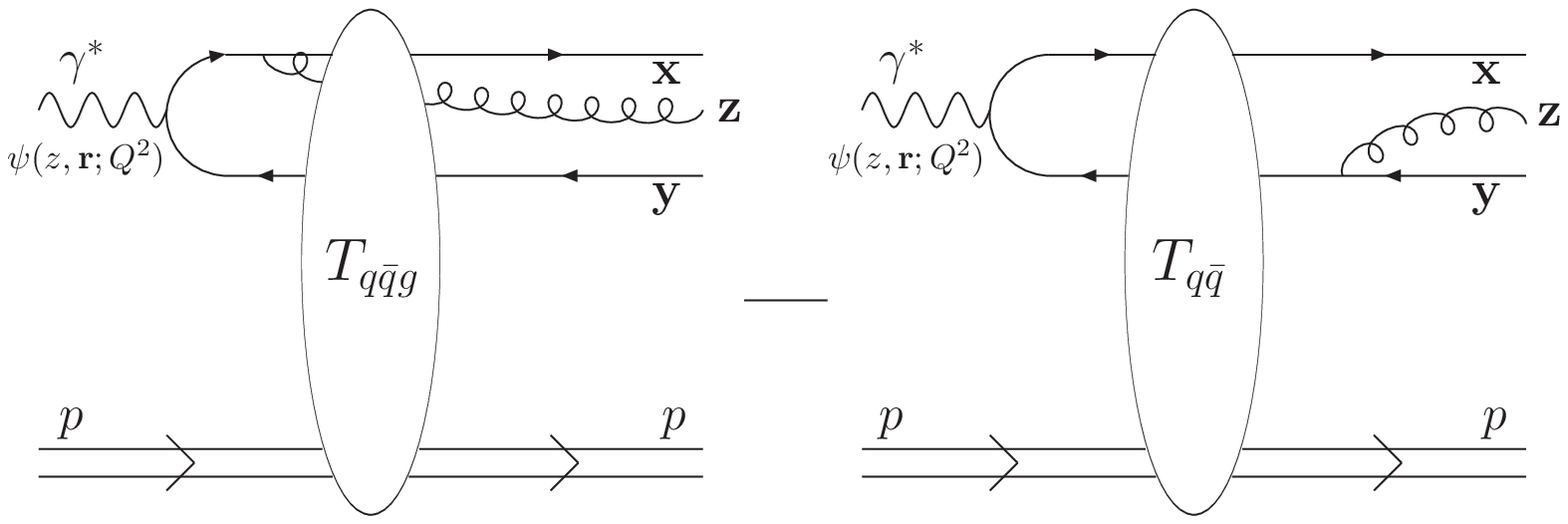}}
\end{minipage}
\caption{The contribution of the $X\!=\!q\bar qg$ final state in diffractive 
$\g^*p\rightarrow Xp$ scattering. Left diagram: at large $Q^2;$ the quark-antiquark transverse distance is much smaller than the quark-gluon transverse distance and an effective $gg$ dipole scatters off the proton. Right diagram: at small $\beta;$ the quark-antiquark-gluon triplet scatters after the gluon emission and the quark-antiquark pair scatters before the gluon emission, with a relative minus sign. In both cases only the amplitude is shown, it has to be squared to obtain the cross-section.}
\end{figure}

\subsection{The small$-\beta$ limit}

At small $\beta,$ the contribution of the $X\!=\!q\bar qg$ final state in diffractive 
$\g^*p\rightarrow Xp$ scattering was computed in many studies
\cite{gbar,gkop,gkov,gmun,gmar}. In coordinate space, denoting $\textbf{x}$ the transverse position of the quark, $\textbf{y}$ that of the antiquark and $\textbf{z}$ that of the gluon, the diffractive scattering is expressed in terms of (see Fig.3b)
\be
\left[T_{q\bar qg}(\textbf{x},\textbf{y},\textbf{z};\xp)
-T_{q\bar q}(\textbf{x},\textbf{y};\xp)\right]^2=S^2(\textbf{b})
\left[N^{(2)}(|\textbf{r}'|Q_s,|\textbf{r}-\textbf{r}'|Q_s,\xp)
-N(|\textbf{r}|Q_s,\xp)\right]^2\ .
\ee
In the left-hand side, the virtual contribution $T_{q\bar q}$ represents the scattering of the quark-antiquark pair, before the gluon emission. Within our model for the impact parameter $\textbf{b}\!=\!(\textbf{x}\!+\!\textbf{y})/2,$ one has
$T_{q\bar q}(\textbf{x},\textbf{y};\xp)=S(\textbf{b})N(|\textbf{r}|Q_s,\xp)$
where the dipole size is naturally $\textbf{r}\!=\!\textbf{x}\!-\!\textbf{y}.$
The real contribution $T_{q\bar qg}$ represents the scattering of the quark-antiquark-gluon triplet, after the gluon emission. In the right-hand side, we factorized the impact parameter profile and wrote $T_{q\bar qg}(\textbf{x},\textbf{y},\textbf{z};\xp)\!=\!
S(\textbf{b})N^{(2)}(|\textbf{r}'|Q_s,|\textbf{r}-\textbf{r}'|Q_s,\xp)$ with 
$\textbf{r}'\!=\!\textbf{x}\!-\!\textbf{z}$ 
(and $\textbf{r}-\textbf{r}'\!=\!\textbf{z}\!-\!\textbf{y}$).

In the context of the BK evolution implemented in \eqref{dipcgc}, the link between 
$N^{(2)}$ and $N$ comes from the fact that the scattering of the $q\bar qg$ triplet is equivalent to the scattering of two dipoles with sizes $\textbf{r}'$ and $\textbf{r}\!-\!\textbf{r}'$ (a dipole emitting a soft gluon is equivalent to a dipole splitting into two dipoles). Therefore our model for $N^{(2)}$ is
\be
N^{(2)}(|\textbf{r}'|Q_s,|\textbf{r}\!-\!\textbf{r}'|Q_s,\xp)=
N(|\textbf{r}'|Q_s,\xp)+N(|\textbf{r}\!-\!\textbf{r}'|Q_s,\xp)-
N(|\textbf{r}'|Q_s,\xp)N(|\textbf{r}\!-\!\textbf{r}'|Q_s,\xp)\ .
\ee

At leading $\ln(1/\beta),$ the contribution of the $q\bar qg$ final state to the transverse diffractive structure function is
\be
\xp\ F_T^{q\bar qg}|_{LL(1/\beta)}(\xp,Q^2)=
\f{C_F\alpha_s Q^2\sigma_0}{8\pi^3\alpha_{em}}
\int_0^\infty rdr \int_0^1 dz\ \Phi_T(z,r;Q^2) A(r,\xp)
\label{qqgbeta}\ee
with
\be
A(|\textbf{r}|,\xp)=\int d^2r' 
\f{\textbf{r}^2}{\textbf{r}'^2(\textbf{r}\!-\!\textbf{r}')^2}
\left[N(|\textbf{r}'|Q_s,\xp)+N(|\textbf{r}\!-\!\textbf{r}'|Q_s,\xp)
-N(|\textbf{r}|Q_s,\xp)
-N(|\textbf{r}'|Q_s,\xp)N(|\textbf{r}\!-\!\textbf{r}'|Q_s,\xp)\right]^2\ .
\label{qqgbfkl}\ee
It is independent of $\beta$ because the structure function 
$F^{D,3}\!\sim\!\beta\ d\sigma^{\g^*p\rightarrow Xp}/d\beta$ picks up the coefficient of 
$\ln(1/\beta)$ in $\sigma^{\g^*p\rightarrow Xp}.$ Also, the overlap function is $\Phi_T$ because in the leading $\ln(1/\beta)$ approximation, the final state mass $M_X$ is fixed only by the soft gluon longitudinal momentum, and therefore transverse sizes are the same in the amplitude and the complex conjugate amplitude.

Note that in \eqref{qqgbeta}, the impact parameter integration $\int d^2b\ S^2(\textbf{b})$ yielded a factor $\sigma^0/4.$ If the $\textbf{b}-$profile was a theta function as assumed in \cite{gmun}, the $q\bar qg$ contribution would be a factor of 2 higher. In what follows, to numerically compute $A(r,\xp),$ we use the clever change of variables introduced in \cite{gmun} that we recall in Appendix A.

\subsection{The model for $\xp F_T^{q\bar qg}$}

\begin{figure}[t]
\begin{center}
\epsfig{file=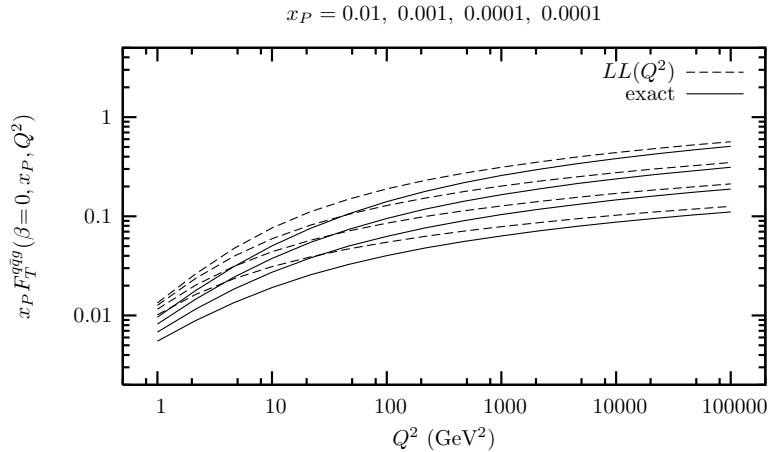,width=10cm}
\caption{The contribution of the $q\bar qg$ final state to the transverse diffractive structure function $F_T^{q\bar qg}$ at $\beta\!=\!0$ as a function of $Q^2.$ The full lines show the exact result $F_T^{q\bar qg}|_{LL(1/\beta)}$ while the dashed lines show the leading $\ln(Q^2)$ result $F_T^{q\bar qg}|_{LL(Q^2)}.$  Different sets of curves are for different values of $\xp=0.01,\ 0.001,\ 0.0001,\ 0.00001,$ from bottom to top. As $Q^2$ increases the two results get closer, but they coincide for only very large values of $Q^2.$}
\end{center}
\end{figure}

The usual approach to implement the $q\bar qg$ contribution is to use formula 
\eqref{qqgQ2}, but as we shall see, this is not correct for small values of $\beta.$
Let us consider the $q\bar qg$ contribution for $\beta\!=\!0.$ By definition, the correct result is $F_T^{q\bar qg}|_{\ln(1\beta)}$ given in formulae 
\eqref{qqgbeta}\!-\!\eqref{qqgbfkl}. By contrast, the small$-\beta$ limit of the leading 
$\ln(Q^2)$ contribution $F_T^{q\bar qg}|_{LL(Q^2)}$ is
\be
\xp\ F_T^{q\bar qg}|_{LL(Q^2)}(\beta\!=\!0,\xp,Q^2)=\f{C_FN_c\alpha_s\sigma_0}{12\pi^4}\sum_f e_f^2\int_0^{Q^2}dk^2\left[\ln\lr{\f{Q^2}{k^2}}\right]
\left|\int_0^\infty\f{dr}{r}J_2(kr)\lr{2N(rQ_s,\xp)\!-\!N^2(rQ_s,\xp)}\right|^2
\label{qqgQ2beta}\ee
where we have used \eqref{qqgQ2} with $K_2(x)\!=\!2/x^2$ for $x\!\to\!0.$ Formula 
\eqref{qqgQ2beta} shows that, after rising as $\beta$ decreases, the diffractive structure function goes to a constant. This constant is different from the correct result 
\eqref{qqgbeta}\!-\!\eqref{qqgbfkl}, except for very large values of $Q^2,$ for which 
$F_T^{q\bar qg}|_{LL(Q^2)}$ is correct by definition. And indeed, if $Q^2\!\gg\!Q_s^2,$ the two formulae coincide to give
\be
\xp\ F_T^{q\bar qg}(\beta\!=\!0,\xp,Q^2\!\gg\!Q_s^2)=
\f{C_FN_c\alpha_sQ_s^2\sigma_0}{6\pi^4}\sum_f e_f^2\ln\lr{\f{Q^2}{Q_s^2}}
\int_0^\infty\f{d\bar{r}}{\bar{r}^3}\left[2N(\bar{r},\xp)-N^2(\bar{r},\xp)\right]^2\ .
\label{twolimits}\ee
This is shown analytically in Appendix B. In Fig.4, we compare formulae 
\eqref{qqgbeta}\!-\!\eqref{qqgbfkl} and \eqref{qqgQ2beta} as a function of $Q^2$ and for different value of $\xp$. One sees that when $Q^2$ increases, the ratio between the two results gets closer to one, but that limit is only reached for very large values of $Q^2$ not shown in the figure. For the values of $Q^2$ in the HERA range, the actual result is smaller than the leading $\ln(Q^2)$ one by a factor of about 0.6.

In order to have the correct $q\bar qg$ contribution for small values of $\beta,$ we shall use the following model:
\be
\xp F_T^{q\bar qg}(\beta,\xp,Q^2)=\xp F_T^{q\bar qg}|_{LL(Q^2)}(\beta,\xp,Q^2)\ 
\f{F_T^{q\bar qg}|_{LL(1/\beta)}(\xp,Q^2)}{F_T^{q\bar qg}|_{LL(Q^2)}(\beta\!=\!0,\xp,Q^2)}
\label{qqgT}\ee
obtained from formulae \eqref{qqgQ2}\!-\!\eqref{intj2k2}, 
\eqref{qqgbeta}\!-\!\eqref{qqgbfkl} and \eqref{qqgQ2beta}. It is such that 
$F_T^{q\bar qg}\!=\!F_T^{q\bar qg}|_{LL(Q^2)}$ at large $Q^2$ and
$F_T^{q\bar qg}\!=\!F_T^{q\bar qg}|_{LL(1/\beta)}$ at small $\beta.$ In the small$-Q^2$ and large$-\beta$ region, the $q\bar qg$ contribution may not be correctly described. However in this case, the diffractive structure function is dominated by the $q\bar q$ component, and the $q\bar qg$ contribution is not relevant.

Finally we point out that our implementation of the $q\bar qg$ contribution is parameter free, the only uncertainty being related to the value of $\alpha_s.$ The average value of $Q^2$ in diffractive measurements at HERA is about 
$10\ \mbox{GeV}^2,$ therefore we choose $\alpha_s\!=\!0.25,$ which corresponds to such a scale.

%%%%%%%%%%%%%%%%%%%%%%%%%%%%%%%%%%%%%%%%%%%%%%%%%%%%%%%%%%%%%%%%%%
%%%%%%%%%%%%%%%%%%%%%%%%   Section 4   %%%%%%%%%%%%%%%%%%%%%%%%%%%
%%%%%%%%%%%%%%%%%%%%%%%%%%%%%%%%%%%%%%%%%%%%%%%%%%%%%%%%%%%%%%%%%%
\section{Description of the HERA data}

The H1 and ZEUS experiments at HERA have measured the diffractive cross section for the process $ep\!\rightarrow\!eXp$, tagging the proton in the final state. After integrating the squared momentum transfer dependence from 
$t_{min}\!=\!-1\ \mbox{GeV}^2$ to $t\!=\!0,$ the data are presented in terms of the reduced cross section $\sigma_r^{D,3}(\beta,\xp,Q^2):$ 
\be
\frac{d^3 \sigma^{ep\rightarrow eXp}}{d\xp\ d\beta\ dQ^2}=\f{4\pi\alpha_{em}^2}{\beta Q^4}
\lr{1-y+\f{y^2}{2}}\sigma_r^{D,3}(\beta,\xp,Q^2)\ ,
\hspace{1cm}
\sigma_r^{D,3}=F_T^{D,3}+\f{2-2y}{2-2y+y^2}\ F_L^{D,3}\ .
\ee
with $y\!=\!Q^2/(sx)$ where $\sqrt{s}\!=\!318\ \mbox{GeV}$ is the total energy in the 
$e\!-\!p$ collision. We shall call the corresponding data sets the LPS \cite{zeuslps} (ZEUS) and FPS \cite{h1fps} (H1) data.

The H1 and ZEUS experiments have also measured the diffractive cross section for the 
process $ep\!\rightarrow\!eXY$, selecting events with a large rapidity gap between the systems $X$ and $Y$ in case of H1 \cite{h1lrg}, and using the so-called $M_X$-method in case of ZEUS \cite{zeusfpc}. $Y$ represents the scattered proton, either intact of in a low-mass excited state, with $M_Y\!<\!1.6\ \mbox{GeV}$ (H1) or $M_Y\!<\!2.3\ \mbox{GeV}$ (ZEUS). The cut on the squared momentum transfer $t$ at the proton vertex, is again
$t\!>\!-1\ \mbox{GeV}^2$ for both experiments. We shall call the corresponding data sets the FPC \cite{zeusfpc} (ZEUS) and LRG \cite{h1lrg} (H1) data. 

Because they include events in which the proton has broken up, the cross-sections measured for the process $ep\!\rightarrow\!eXY$ are larger than the one measured for the process $ep\!\rightarrow\!eXp.$ Also, because H1 and ZEUS measurements are performed with different $M_Y$ cuts, the ZEUS cross-section is bigger than the H1 cross-section, for which the proton-dissociative events are more reduced. However, within the kinematical ranges of the measurements, it seems that the differences are constant factors: the FPC and LRG data points can be converted to the FPS-LPS ones by dividing the cross-sections by $1.45$ and $1.23$ respectively \cite{zeusfpc,h1lrg}. Note that it is the FPS-LPS data that correspond to our definition of diffractive events and to our formulae, as the proton should truelly escape the collision intact.

In our model, the reduced cross section $\sigma_r^{D,3}(\beta,\xp,Q^2)$ is given in terms of the diffractive structure functions by
\be
\xp\sigma_r^{D,3}=\xp F_T^{q\bar q}+\xp F_T^{q\bar qg}
+\f{2-2y}{2-2y+y^2}\ \xp F_L^{q\bar q}\ .
\ee
Using formula \eqref{qqT}, \eqref{qqL}, \eqref{qqgT}, and the dipole scattering amplitude 
\eqref{dipcgc}, we obtain a parameter free calculation for $\xp\sigma_r^{D,3}$ that we can compare to the data. Diffractive DIS measurements are sensitive to the saturation regime of QCD only for small values of $\xp,$ therefore we shall only consider experimental data which feature 
$\xp\!<\!10^{-2}$ in our comparisons. Note that we do not include any 
$q\bar qg$ contribution to the longitudinal structure function: for small values of $\beta$ it could be sizeable, but for kinematical reasons small $\beta$ is associated with $y$ close to $1,$ in which case $F_L^{D,3}$ does not contribute to
$\sigma_r^{D,3}.$

\begin{figure}[ht]
\begin{center}
\epsfig{file=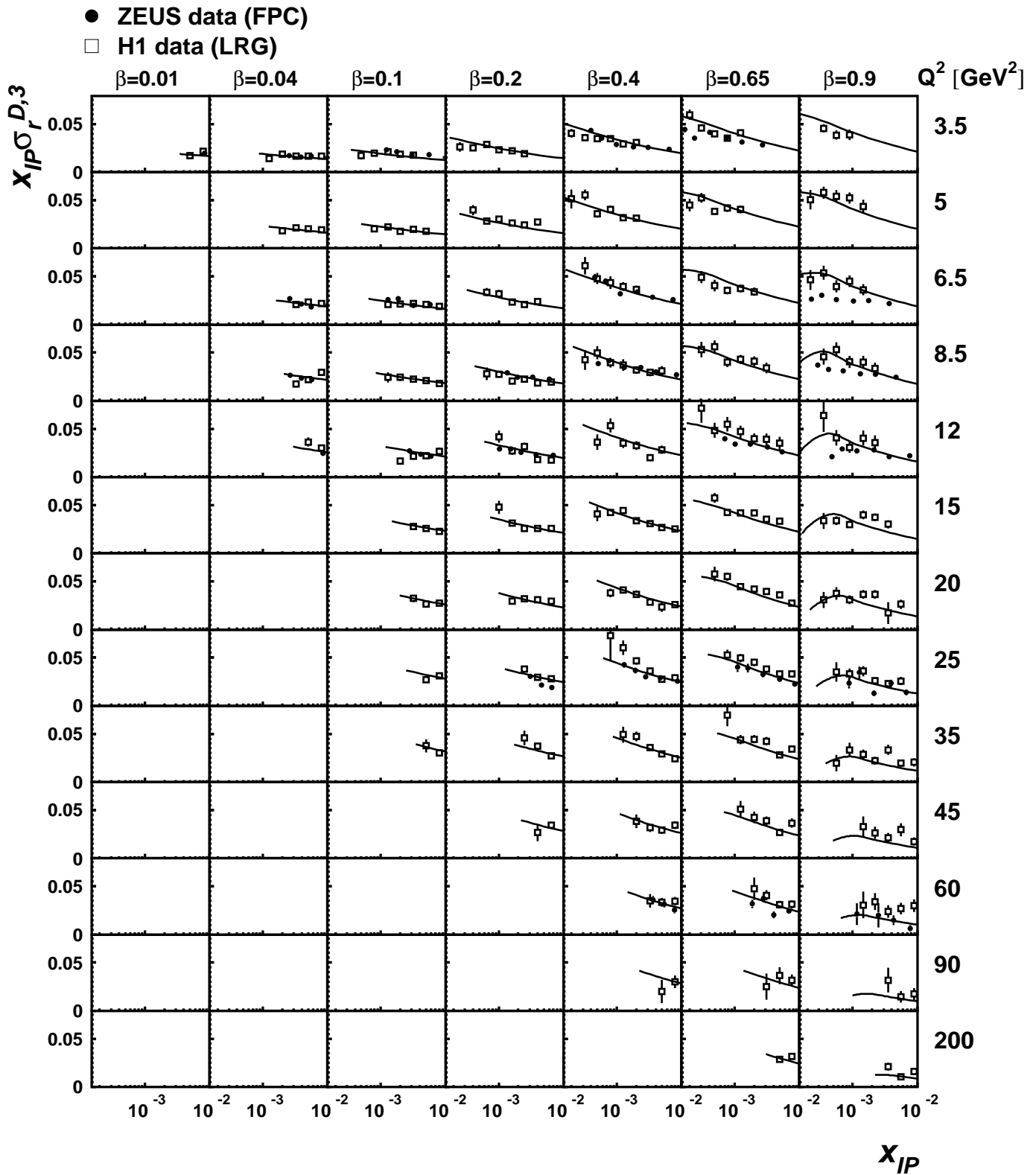,width=\textwidth}
\caption{$\xp \sigma_r^{D,3}(\beta,\xp,Q^2)$ as a function of $\xp$ for different values of $\beta$ and $Q^2.$ The H1 (LRG) and ZEUS (FPC) diffractive data are compared to the predictions of our model, and the $\xp$ range is restricted to 
$\xp\!<\!0.01.$ In this figure, the H1 data are unchanged and it is our predictions which are multiplied by the factor 1.23. The ZEUS data are multiplied by $0.85$ in order to convert them to the H1 $M_Y$ range and the bins centers have been shifted to the H1 values using a parametrisation given in \cite{zeusfpc}. Only the statistical part of the uncertainty is shown for the data points on this plot. 
The shape of the curves in the $\beta\!=\!0.9$ bins is due to the fact that the contribution of $F_L^{D,3}$ to the reduced cross-section is important for large values of $\beta.$}
\end{center}
\end{figure}

\clearpage

To estimate the quality of our description, we performed the following $\chi^2$ computations, adding statistical and systematic uncertainties in quadrature. Within the LPS+FPS ($ep\!\rightarrow\!eXp$) data sets, 76 points pass the $\xp\!<\!0.01$ cut and we obtain $\chi^2/\mbox{points}=0.80.$ When comparing to the 4 data sets, with the proper renormalizations for the FPC and LRG ($ep\!\rightarrow\!eXY$) measurements, 343 points pass the $\xp\!<\!0.01$ cut and we obtain 
$\chi^2/\mbox{points}=1.28.$ This is a quite good description, considering our predictions are parameter free. As an illustration, Fig.5 displays a comparison of our predictions with the FPC+LRG data. We also checked the agreement with the charm contribution to $\sigma_r^{D,3}$ using the few points available \cite{f2dc}, one obtains $\chi^2/\mbox{points}=0.68.$

\begin{figure}[t]
\begin{center}
\epsfig{file=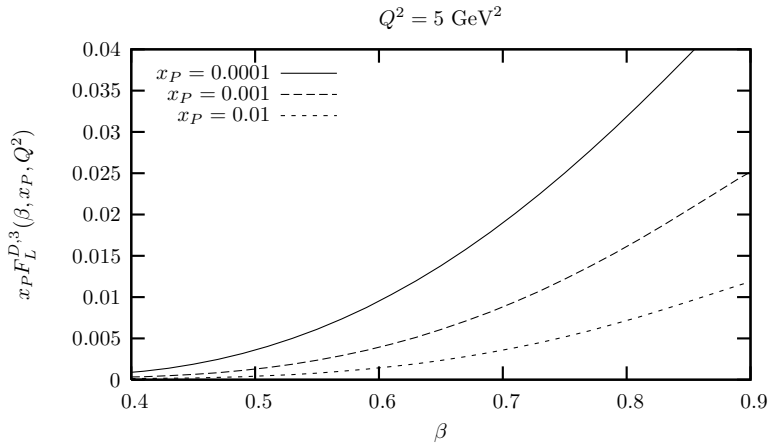,width=10cm}
\caption{Predictions for the longitudinal diffractive structure function.
$\xp F_L^{D,3}(\beta,\xp,Q^2)$ is plotted as a function of $\beta$ for 
$Q^2=5\ \mbox{GeV}^2$ and for different values of $\xp=0.01,\ 0.001,\ 0.0001.$ With our parametrization, only the $q\bar q$ final state contributes.}
\end{center}
\end{figure}

Note that, if the running coupling $\alpha_s(Q^2)$ is used in the $q\bar qg$ contribution (instead of imposing $\alpha_s\!=\!0.25$), the description of 
$\sigma_r^{D,3}$ is also very good with similar values of $\chi^2/\mbox{points}:$ 0.63 when comparing to the LPS and FPS data, and 1.20 when comparing to the four data sets. In this case, the additional cut $Q^2\!>\!1\ \mbox{GeV}^2$ is used in order to keep the coupling reasonably small (this only removes 15 of the LPS points). We also noticed that it is possible to obtain a description of equal quality without the correction 
$F_T^{q\bar qg}|_{LL(1/\beta)}/F_T^{q\bar qg}|_{LL(Q^2)}$ in the $q\bar qg$ contribution 
\eqref{qqgT}, if one imposes an unphysically small value for the coupling: $\alpha_s=0.15.$

Finally, Fig.6 shows predictions for the longitudinal diffractive structure function $F_L^{D,3}(\beta,\xp,Q^2),$ which in our approach in obtained from the $q\bar q$ contribution 
$F_L^{q\bar q}$ (see formula \eqref{qqL}).

%%%%%%%%%%%%%%%%%%%%%%%%%%%%%%%%%%%%%%%%%%%%%%%%%%%%%%%%%%%%%%%%%%
%%%%%%%%%%%%%%%%%%%%%%%   Conclusion   %%%%%%%%%%%%%%%%%%%%%%%%%%%
%%%%%%%%%%%%%%%%%%%%%%%%%%%%%%%%%%%%%%%%%%%%%%%%%%%%%%%%%%%%%%%%%%
\section{Conclusions}

We presented a new description of HERA diffractive deep inelastic scattering data. It uses the parametrization for the dipole scattering amplitude obtained in 
\cite{greg}. This is an extention of the IIM saturation model which contains heavy-quark contributions. Contrary to previous studies, it features a saturation scale $Q_s(x)$ that stays above 1 GeV for $x\!=\!10^{-5},$ rather than dropping by a factor of 2 after including heavy-quark contributions. Our description of the data is parameter-free and, in the regime $\xp\!<\!0.01$ in which saturation is expected to be relevant, it features values of $\chi^2/\mbox{points}$ of order 1.

Let us recall the improvements that our model brings with respect to previous approaches.
\begin{itemize}
\item Instead of considering the dipole cross-section $\sigma_0$ and the diffractive slope $B_D$ as independent quantities, we use the relation \eqref{consipd}. It results from a consistent treament of the impact parameter dependence of the dipole scattering amplitude.

\item In the large$-Q^2$ limit, the $q\bar qg$ contribution to the diffractive final state is described by a $gg$ dipole. It is related in terms the $q\bar q$ dipole in such a way that it is consistent with the BK evolution implemented in \eqref{dipcgc}. 

\item The $q\bar qg$ contribution to the diffractive final state is modeled in such a way that both the large$-Q^2$ and small$-\beta$ limits are implemented \eqref{qqgT}. This allows to have a good descrition of the data with a consistent value of 
$\alpha_s\!=\!0.25.$

\item Our predictions include correclty the contribution of the longitudinal diffractive structure function $F_L^{D,3}:$ we predict $\sigma_r^{D,3},$ not 
$F_2^{D,3}\!=\!F_T^{D,3}\!+\!F_L^{D,3}.$

\item When comparing our predictions to the experimental data, the comparison is made with the $ep\!\rightarrow\!eXp$ data unchanged and the $ep\!\rightarrow\!eXY$ data renormalized, and not the opposite. Our definition of diffractive events (and our formulae) are such that the proton truelly escape the collision intact. 
\end{itemize}

Of all the possible descriptions of diffractive DIS data (for a global analysis, see 
\cite{globdiff}), the dipole picture is the one which is adapted to study the physics of parton saturation. Having a consistent saturation model to describe hard diffraction in $e\!-\!p$ scattering represent a good foundation for further works. On the phenomelogical side, studying hard diffraction in $e\!-\!A$ becomes of interest \cite{ddisateic}, as it is an ideal process to investigate the saturation regime of QCD that could be explored at a future electron-ion collider. On the theoretical side, many new developements improved our understanding of the QCD 
non-linear evolution, and their consequences are to be investigated
\cite{diffscal,difftheo}.

%%%%%%%%%%%%%%%%%%%%%%%%%%%%%%%%%%%%%%%%%%%%%%%%%%%%%%%%%%%%%%%%%%
%%%%%%%%%%%%%%%%%%%%%   Acknowledgments   %%%%%%%%%%%%%%%%%%%%%%%%
%%%%%%%%%%%%%%%%%%%%%%%%%%%%%%%%%%%%%%%%%%%%%%%%%%%%%%%%%%%%%%%%%%
\begin{acknowledgments}

I am grateful to Gr\'egory Soyez for sharing his results \cite{greg} on the heavy-quark improved IIM saturation model while they were still preliminary. I am also grateful to Laurent Schoeffel for helping collecting the HERA data on inclusive diffraction, for helping building up Fig.5, and for useful clarifications concerning the experimental measurements. I would like to thank them as well as Tuomas Lappi and Raju Venugopalan for insightful discussions concerning several aspects of this work. I also acknowledge Krzyzstof Golec-Biernat and Robi Peschanski for useful suggestions. I thank the Galileo Galilei Institute for Theoretical Physics for hospitality and the INFN for partial support during the completion of part of this work. This research was supported in part by RIKEN, Brookhaven National Laboratory and the U.S. Department of Energy [DE-AC02-98CH10886].
 
\end{acknowledgments}
%%%%%%%%%%%%%%%%%%%%%%%%%%%%%%%%%%%%%%%%%%%%%%%%%%%%%%%%%%%%%%%%%%
%%%%%%%%%%%%%%%%%%%%%%%%%   Appendix   %%%%%%%%%%%%%%%%%%%%%%%%%%%
%%%%%%%%%%%%%%%%%%%%%%%%%%%%%%%%%%%%%%%%%%%%%%%%%%%%%%%%%%%%%%%%%%
\appendix
\section{Numerical computation of $A(|\textbf{r}|,\xp)$}

In this Appendix, following \cite{gmun}, we show how to transform the expression (see formula \eqref{qqgbfkl})
\be
A(|\textbf{r}|,\xp)=\int d^2r' 
\f{\textbf{r}^2}{\textbf{r}'^2(\textbf{r}\!-\!\textbf{r}')^2}
\left[N(|\textbf{r}'|Q_s,\xp)+N(|\textbf{r}\!-\!\textbf{r}'|Q_s,\xp)
-N(|\textbf{r}|Q_s,\xp)
-N(|\textbf{r}'|Q_s,\xp)N(|\textbf{r}\!-\!\textbf{r}'|Q_s,\xp)\right]^2
\ee
in order to estimate it numerically. Writing the two-dimensional integration in the complex plane and introducing $S=1-N,$ we obtain (with $r\!=\!|\textbf{r}|$)
\be
A(r,\xp)=\int \f{dzd\bar z}{2|z|^2|1\!-\!z|^2}
\left[S(|z|rQ_s,\xp)S(|1\!-\!z|rQ_s,\xp)-S(rQ_s,\xp)\right]^2\ .
\ee
We then follow the following procedure.
\begin{itemize}
\item $|z|$ and $|1-z|$ are invariant by symmetry with respect to the real axis so one can multiply the integral by 2 and restrict ourselves to the upper part of the complex plane.
\item For $|z|\leq 1,$ let us change the variables into $u=|z|\in[0,1]$ and 
$v=\f{|1-z|+|z|-1}{2|z|}\in[0,1],$ this implies
\be dzd\bar z=\f{8|z|^2|1\!-\!z|dudv}{|z\!-\!\bar z|}\hspace{1cm}
|1\!-\!z||z\!-\!\bar z|=4u(1\!-\!u\!+\!2uv)\sqrt{v(1\!-\!v)(1\!+\!uv)(1\!-\!u\!+\!uv)}\ .\ee
\item For $|z|\geq 1,$ let us change the variables into $u=1/|z|\in[0,1]$ and 
$v=\f{|1-z|-|z|+1}{2}\in[0,1],$ this implies
\be dzd\bar z=\f{8|z|^3|1\!-\!z|dudv}{|z\!-\!\bar z|}\hspace{1cm}
\f{|1\!-\!z|}{|z|}|z\!-\!\bar z|=4(1/u\!-\!1\!+\!2v)
\sqrt{v(1\!-\!v)(1\!+\!uv)(1\!-\!u\!+\!uv)}\ .\ee
\end{itemize}
One gets
\bea
A(r,\xp)=\int_0^1\int_0^1 \f{2dudv}{u(1\!-\!u\!+\!2uv)
\sqrt{v(1\!-\!v)(1\!+\!uv)(1\!-\!u\!+\!uv)}}\hspace{3cm}
\nn\\\left\{u^2[S(rQ_s/u,\xp)S((1/u\!-\!1\!+\!2v)rQ_s,\xp)-S(rQ_s,\xp)]^2
\right.\nn\\\left.+[S(urQ_s,\xp)S((1\!-\!u\!+\!2uv)rQ_s,\xp)-S(rQ_s,\xp)]^2\right\}
\eea
which is easy to evaluate numerically, as this features only integrable singularities.

\section{The small$-\beta$ and large$-Q^2$ limit of $\xp F_T^{q\bar qg}$}

In this Appendix, we show that the small$-\beta$ limit of the leading $\ln(Q^2)$ result $F_T^{q\bar qg}|_{LL(Q^2)}$ and the large$-Q^2$ limit of the leading 
$\ln(1/\beta)$ result $F_T^{q\bar qg}|_{LL(1/\beta)}$ coincide. The first case has been derived in the text and we obtained formula \eqref{twolimits}. We now show how, for $Q^2\!\gg\!Q_s^2,$ formulae \eqref{qqgbeta}\!-\!\eqref{qqgbfkl} give the same result.

The starting point is the transverse overlap function
\be
\Phi_T(z,r;Q^2)=\frac{\alpha_{em}N_c}{2\pi^2}\sum_f e_f^2
\lr{[z^2+(1\!-\!z)^2]\varepsilon_f^2 K^2_1(r\varepsilon_f)
+m_f^2K^2_0(r\varepsilon_f)}\ ,\label{phit}
\ee
where one can neglect quark masses with respect to $Q^2.$ The $z$ integration of 
\eqref{phit} can be done in the two limits $rQ\!\ll\!1$ and 
$rQ\!\gg\!1:$ using the Mellin representation of $K^2_1(x),$ one gets
\bea
\int_0^1 dz [z^2+(1\!-\!z)^2]z(1\!-\!z)K^2_1\left[\rho\sqrt{z(1\!-\!z)}\right]&=&
\sqrt{\pi}\int_{c-i\infty}^{c+i\infty}\f{d\g}{2i\pi}\rho^{-2\g}
\f{\Gamma(\g\!-\!1)\Gamma(\g)\Gamma(\g\!+\!1)\Gamma(2\!-\!\g)\Gamma(4\!-\!\g)}
{\Gamma(6\!-\!2\g)\Gamma(\g\!+\!1/2)}\nn\\&&1\!<\!\mbox{Re}(c)\!<\!2
\nn\\\nn\\&=&\f23\left\{\begin{array}{lll}\D4/\rho^4\mbox{ for } \rho\gg1
\\\\\D1/\rho^2\mbox{ for } \rho\ll1\end{array}\right.\ .
\eea
To obtain the second equality, we used the fact that in the $\rho\!\gg\!1$ case 
($\rho\!\ll\!1$ case), the dominant contribution to the $\g$ integration comes from the single pole at $\g\!=\!2$ (at $\g\!=\!1$). One can then write
\be
\int_0^\infty rdr \int_0^1 dz\ \Phi_T(z,r;Q^2) f(r)=
\f{\alpha_{em}N_c}{3\pi^2}\sum_f e_f^2\left[\int_0^{2/Q}\f{dr}{r}f(r)
+4\f1{Q^2}\int_{2/Q}^\infty\f{dr}{r^3}f(r)\right]\ .
\label{estpht}\ee
Note that in \cite{gmun}, a similar estimation was obtained. By contrast, the replacement 
$K_1(\rho)\!\to\!\Theta(1\!-\!\rho)/\rho$ was used and as a result, the normalization factors were not under control.

To complete the calculation, we need to compute $f(r)\!=\!A(r,\xp),$ in the two limits 
$rQ_s\!\ll\!1$ and $rQ_s\!\gg\!1$ (see \cite{gmun}):
\bea
A(r,\xp)&=&2\pi r^2Q_s^2\int\f{d\bar{r}}{\bar{r}^3}
\left[2N(\bar{r},\xp)-N^2(\bar{r},\xp)\right]^2\mbox{ for }rQ_s\!\ll\!1\\
A(r,\xp)&=&2\pi\ln(r^2Q_s^2) [1-N(rQ_s,\xp)]^2 \mbox{ for }rQ_s\!\gg\!1\ .
\eea
The first line is obtained from configurations in which the $q\bar q$ pair is small and is well separated from the gluon. When using those results in \eqref{estpht} by dividing the $r$ integration region in three domains, one sees that the dominant contribution comes from the region $r\in[2/Q,1/Q_s].$ It is enhanced by the collinear factor 
$\ln(Q^2/Q_s^2):$
\be
\xp\ F_T^{q\bar q g}(\beta\!=\!0,Q^2\!\gg\!Q^2_s)=
\f{C_FN_c\alpha_sQ_s^2\sigma_0}{6\pi^4}
\sum_f e_f^2\ln\lr{\f{Q^2}{Q_s^2}}\int_0^\infty\f{d\bar{r}}{\bar{r}^3} 
\left[2N(\bar{r},\xp)-N^2(\bar{r},\xp)\right]^2\ .
\ee
This formula is identical to formula \eqref{twolimits}.

%%%%%%%%%%%%%%%%%%%%%%%%%%%%%%%%%%%%%%%%%%%%%%%%%%%%%%%%%%%%%%%%%%
%%%%%%%%%%%%%%%%%%%%%%   Bibliography   %%%%%%%%%%%%%%%%%%%%%%%%%%
%%%%%%%%%%%%%%%%%%%%%%%%%%%%%%%%%%%%%%%%%%%%%%%%%%%%%%%%%%%%%%%%%%

\end{document}